\documentclass[prd,nofootinbib,superscriptaddress,showpacs,showkeys]{revtex4}
\usepackage{graphicx,epsfig}
\usepackage{amsfonts,amsmath,amssymb,amsthm,amscd}

\begin{document}

\title{Boost modes for a massive fermion field}

\author{E. G. Gelfer}
\author{A.M. Fedotov}
\author{V.D. Mur}
\author{N.B. Narozhny}
\affiliation{National Research Nuclear University MEPhI, 115409
Moscow, Russia}

\begin{abstract}

We have shown that Wightman function of a free quantum field
generates any complete set of solutions of relativistic wave
equations. Using this approach we have constructed the complete set
of solutions to 2d Dirac equation consisting of eigenfunctions of
the generator of Lorentz rotations (boost operator). It is shown
that at the surface of the light cone the boost modes for a fermion
field contain $\delta$-function of a complex argument. Due to the
presence of such singularity exclusion even of a single mode with an
arbitrary value of the boost quantum number makes the set of boost
modes incomplete.
\end{abstract}

 \maketitle

\section{Introduction}

It is well known that quantization procedure in a quantum field
theory implies expansion of the field operator in terms of a
complete set of modes which are solutions of the corresponding
(Dirac or Klein-Fock-Gordon) classical field equation. Therefore finding
exact solutions for these equations is one of the central points of
the quantum field theory in the presence of a classical background.

One of the most powerful instruments for solving partial
differential equations is based on symmetry properties of the
physical system described by the equation. According to the
Noether's theorem \cite{vved} any differentiable symmetry of the
action of a physical system has a corresponding conservation law.
One can always choose a set of variables in such a way that a change
in one of them corresponds to the symmetry transformation. Since the
generator of the symmetry transformation commutes with the
differential operator of the equation, the variable corresponding to
it can be separated. If there are enough symmetries for the
equation, such that their generators commute with each other, the
problem of finding the solution can be reduced to solving, generally
speaking, of a second order ordinary equation. The set of such
solutions is labeled by the eigenvalues of the generators and is
complete.

All most important and widely used solutions for relativistic
quantum equations in the presence of classical external fields were
obtained using this method. We mean the Coulomb field, see, e.g.,
\cite{ahiezer}, the constant magnetic \cite{ahiezer} and electric
\cite{nik,TMF} fields, the field of a plane electromagnetic wave
\cite{volkov},  and some other fields of more sophisticated
configurations \cite{Redm,TMF}. In all these cases the solutions
were eigenfunctions of linear combinations of generators of time and
spatial translations or spatial rotations, and were labeled by
values of components of either 4-momentum, or angular momentum
respectively. However, the Poincar\'{e} group, the group of
isometries of Minkowski spacetime (MS), includes also Lorentz
rotations (boosts). The boost symmetry was almost never used for
field quantization. Certainly, this is because the boost generator
does not commute with the Hamiltonian, and thus the boost quantum
number and energy cannot be parts of any complete set of observables
simultaneously. Nevertheless, the boost modes can be very useful,
especially for the quantum field theory in a curved space, or if the
symmetry of MS with respect to time and/or space translations is
broken in the presence of a classical background. In these cases the
boost symmetry can appear to be the sole symmetry for the quantum
field. Dilaton gravity in two dimensions \cite{dilgr}, as well as
Schwarzschild geometry \cite{Schw}, are the examples. Therefore
analysis of the properties of boost modes is important.

For the first time the boost modes for a free massive scalar field
were discussed in Ref.~\cite{unruh} by W. Unruh, though their
explicit form was not specified. Then some properties of scalar
boost modes were studied in Refs.~\cite{nikrit,gerlach,narozhny}. In
particular, the remarkable properties of boost modes in 2d MS were
ascertained in Ref.~\cite{narozhny}. It was shown that (i) the zero
boost mode of a free massive scalar field coincides up to a trivial
constant factor with the positive-frequency Whightman function, and
(ii) the boost modes considered as functions of the boost quantum
number possess the Dirac $\delta$-function singularity at the
surface of the light cone. The properties of the boost modes for the
case of a free massless fermion field appeared to be even more
interesting. It was shown in Ref.~\cite{jetpl} that boost modes of
two-dimensional massless fermions on a light cone are expressed in
terms of the delta function of a complex argument. Boost modes of a
massive fermion field have been first considered in
Refs.~\cite{muller,mullerbook}, see also Ref.~\cite{McMah}. However
the results of Refs.~\cite{muller,mullerbook,McMah} contain some
discrepancies which we discuss in Sec. III of the present paper.

In this paper we consider boost modes for a free massive fermion
field. To construct them we start from the Wightman functions. The
positive and negative frequency Wightman functions for free fields
are explicitly determined (with accuracy to a constant factor) by
Lorentz and translational invariance of the theory \cite{jost}. And
vice versa, once the Wightman functions are known, one can use those
symmetries to construct any complete set of positive and negative
frequency solutions of Klein-Fock-Gordon (KFG) or Dirac equations.
In fact, this is the direct consequence of the Wightman
reconstruction theorem \cite{streater,bogolubov}.

Since the Lorentz rotation singles out a two-dimensional plane in
MS, we will discuss the specific properties of boost modes by
examples of free 2d KFG and Dirac equations. In the next section we
will realize the above formulated approach by the example of a
single-component massive neutral field. It is worth noting that, as
opposed to the case of plane wave modes, analytic properties of
boost modes change dramatically when one passes to multi-component
fields which are considered in Sec. III. The discussion of the
results and conclusions are given in Sec.~IV.

\section{Bosons}

In this section we will consider the case of a free neutral massive
scalar field in two-dimensional MS. We will start from two-point
Wightman function \cite{streater} which for a free field theory
coincides with the positive-frequency part of the commutator of two
field operators (Pauli-Jordan function in 4d theory of a scalar
field). Positive-frequency Wightman function
$\Delta^{(+)}(x),\,x=(t,z),$ for a massive scalar field satisfies
KFG equation\footnote{The natural units  $\hbar =c=1$ are used throughout this paper.}
\begin{equation}\label{kg}
    \mathcal{K}\Delta^{(+)}(x)=0,\quad\mathcal{K}=\partial^2_t-\partial^2_z+m^2,
\end{equation}
contains only positive frequencies and is invariant with respect to
Lorentz rotations (boosts)
\begin{equation}\label{boostgen}
  \mathcal{B_K}\Delta^{(+)}(x)=0,\quad\mathcal{B_K}=i(z\partial_t+t\partial_z).
\end{equation}
These conditions determine $\Delta^{(+)}(x)$ accurate within a
constant factor. Indeed, any positive-frequency solution of KFG
equation can be written as follows
\begin{equation}\label{pos_fr}
\Phi^{(+)}(x)=\int d^2p\,\phi(p)\delta(p^2-m^2)\theta(p^0)e^{-ipx},
\end{equation}
where $\phi(p)$ is a certain function of 2-vector $p=(p^0,p^1)$,
$\theta(p^0)$ is the Heaviside step function. After the change of
variables
$$p^0=\mu\cosh q,\, p^1=\mu\sinh q,$$ where $q$ is rapidity, and
integration over $\mu$ we obtain the equation
\begin{equation}\label{pos_fr1}
\Phi^{(+)}(x)=\frac{1}{2}\int\limits_{-\infty}^{\infty} dq\,\phi(m\cosh q,m\sinh q)
e^{-im(t\cosh q-z\sinh q)}.
\end{equation}
It can be easily seen that condition (\ref{boostgen}) is satisfied
only if $\phi(q)=const$. Choosing $\phi(q)=i/2\pi$ we arrive to the
standard representation for $\Delta^{(+)}(x)$, compare, e.g.,
\cite{vved,narozhny}
\begin{equation}\label{scwf1}
\Delta^{(+)}(x)=\frac{i}{4\pi}\int\limits_{-\infty}^\infty
dqe^{-im(t\cosh q-z\sinh q)}.
\end{equation}
It is assumed in (\ref{scwf1}) that an infinitely small negative
imaginary part is added to $t$ \cite{narozhny}. Another
representation for $\Delta^{(+)}(x)$ reads, see, e.g.,
\cite{wightman},
\begin{equation}\label{scwf}
\Delta^{(+)}(x)=\frac{i}{4\pi}\int\limits_{-\infty}^\infty\frac{dp}
{\varepsilon_p}e^{-i\varepsilon_pt+ipz},\quad
\varepsilon_p=\sqrt{p^2+m^2},
\end{equation}
from now on we shall omit the index of the spatial component of the 2-vector $p$.

According to the Wightman reconstruction theorem \cite{streater} the
two-point Wightman function uniquely determines the quantum theory
of a free field. Particularly, it allows to reconstruct any
orthonormalized and complete set of solutions for the field
equation. Indeed, it follows from the translational invariance of
the theory that Wightman function with an arbitrarily shifted
argument
\begin{equation}\label{wf_sh}
\Delta^{(+)}(x-u), \quad u=\{u^0,u^1\}
\end{equation}
satisfies KFG equation. Functions (\ref{wf_sh}) constitute an
overcomplete set of modes since they are labeled by two independent
parameters $u^0,u^1$. Anyway, any complete set of
positive-frequency solutions $F_a(x)$
\begin{equation}\label{compl}
\int da F_a(x')F_a^*(x'')=-i\Delta^{(+)}(x'-x''),
\end{equation}
orthonormalized by the condition
\begin{equation}\label{scnorm}
  (F_a,F_{a'})\equiv i\int\limits_{-\infty}^\infty
  F_a^*(x)\overset{\leftrightarrow}{\partial}_tF_{a'}(x)dz=\delta(a-a')
\end{equation}
can be represented as
\begin{equation}\label{razl}
 F_a(x)=\int d^2u f_a(u)\Delta^{(+)}(x-u).
\end{equation}
However, since the set of solutions (\ref{wf_sh}) is overcomplete,
the coefficient functions $f_a(u)$ cannot be determined
uniquely. To make the choice of $f_a(u)$ unique one should
impose a restriction on $u$ which can be chosen for reasons of
symmetry.

First, we will illustrate this procedure for the trivial case of
plane waves $\Theta_p(x)$ which are eigenfunctions of the generator
of spatial translations
\begin{equation}\label{planesimm}
  -i\partial_z\Theta_p(x)=p\Theta_p(x),
\end{equation}
In this case it is convenient to limit the family of
$\Delta^{(+)}(x-u)$ only by functions spatially shifted with
respect to each other. This means that coefficient functions
$f_p(u)$ have the form
$f_p(u)=\delta(u^0)\vartheta_p(u^1)$, so that
\begin{equation}\label{planerazl}
  \Theta_p(x)=\int\limits_{-\infty}^\infty du^1
  \vartheta_p(u^1)\Delta^{(+)}(t,z-u^1).
\end{equation}
Let us substitute this expansion into Eq.~(\ref{planesimm}). Taking
into account that Wightman function $\Delta^{(+)}(t,z)$
exponentially tends to zero when $|z|\rightarrow\infty$, we can
integrate the left-hand side of the resulting equation by part and
arrive to the following equation for $\vartheta_p(u^1)$
$$-i\vartheta'_p(u^1)=p\,\vartheta_p(u^1)\,,$$
so that $\vartheta_p(u^1)=const\cdot e^{ip\,u^1}$.
Substituting now $\vartheta_p(u^1)$ of that form into
(\ref{planerazl}) and using (\ref{scwf}) we finally obtain after
integration over $u^1$
\begin{equation}\label{scplane}
  \Theta_p(x)=\frac{1}{\sqrt{4\pi\varepsilon_p}}\,e^{-i\varepsilon_p t+ipz}.
\end{equation}
The normalization constant here was determined from the condition
(\ref{scnorm}).

Functions $\Theta_p(x)$ satisfy the relation
\begin{equation}\label{compl_p}
\int\limits_{-\infty}^{\infty}dp\,\Theta_p(x')\Theta^*_p(x'')=
-i\Delta^{(+)}(x'-x''),
\end{equation}
and thus the set (\ref{scplane}) is complete.

We are interested in the set of boost modes $\Psi_\kappa^{(+)}(x)$
which are eigenfunctions of boost operator $\mathcal{B_K}$
\begin{equation}\label{boostgen1}
  \mathcal{B_K}\Psi_\kappa^{(+)}(x)=\kappa\Psi_\kappa^{(+)}(x)\,.
\end{equation}
It is worth noting from the very beginning that Wightman function
$\Delta^{(+)}(x)$ is a zero mode $\Psi_0^{(+)}(x)$ of this set as it
follows from Eqs.~(\ref{boostgen1}) and (\ref{boostgen}).

To obtain the boost modes we will confine $u$ in (\ref{razl}) to one
of the orbits of the restricted Lorentz group
\begin{equation}\label{orbits}
u^2={u^0}^2-{u^1}^2=\pm v^2\,, \quad
v=const\,,
\end{equation}
so that $f(u)\sim\delta(u^2\mp v^2)$. Choosing the
upper sign in (\ref{orbits}) for definiteness, we put
\begin{equation}\label{alph}
u\equiv u_q=(v\cosh q,v\sinh q),
\end{equation}
and rewrite Eq.~(\ref{razl}) in the form
\begin{equation}\label{boostrazl}
  \Psi_\kappa^{(+)}(x)=\int\limits_{-\infty}^\infty dq
  \zeta_\kappa(q)\Delta^{(+)}(x-u_q).
\end{equation}
Substituting (\ref{boostrazl}) into Eq.~(\ref{boostgen1}) and using
the relation
\begin{equation}\label{bwf}
\mathcal{B_K}\Delta^{(+)}(x-u_q)=-i\frac{\partial}{\partial
q} \Delta^{(+)}(x-u_q)\,,
\end{equation}
which can be straightforwardly
obtained with the help of representation (\ref{scwf1}), we get the
following equation for $\zeta_\kappa(q)$
$$i\frac{\partial \zeta_{\kappa}(q)}{\partial q}=\kappa
\zeta_{\kappa}(q)\,.$$ Hence, $ \zeta_{\kappa}(q)=const\,e^{-i\kappa q}$,
and we finally get from (\ref{boostrazl}) with due regard for
condition (\ref{scnorm}) the following representation for the
positive frequency boost modes
\begin{equation}\label{scboostm}
  \Psi_\kappa^{(+)}(x)=\frac{1}{2^{3/2}\pi}\int\limits_{-\infty}^\infty dq
  e^{-im(t\cosh q-z\sinh q)-i\kappa q},
\end{equation}
which was earlier obtained in Refs.~\cite{nikrit,gerlach,narozhny} by
another method. The negative frequency boost modes
$\Psi_\kappa^{(-)}(x)$ are defined \cite{narozhny} as
\begin{equation}\label{scboostm1}
\Psi_\kappa^{(-)}(x)={\Psi_{-\kappa}^{(+)}}^*(x)\,.
\end{equation}
Boost modes (\ref{scboostm}), (\ref{scboostm1}) are distributions
with respect to both $x=(t,z)$ and spectral parameter $\kappa$. They
are defined on the class of smooth functions of rapid enough
descent.

The modes (\ref{scboostm}),(\ref{scboostm1}) constitute a complete
set since they satisfy the condition
\begin{equation}\label{boost_compl}
\int\limits_{-\infty}^\infty
  d\kappa\Psi_\kappa^{(\pm)}(x'){\Psi_\kappa^{(\pm)}}^*(x'')=\mp i\Delta^{(\pm)}(x'-x''),
\end{equation}
and hence can be used as a basis for quantization of a neutral
scalar field, see Ref.~\cite{narozhny}. Since the sign of a particle
energy is Lorentz invariant, the vacuum states in the boost and
plane-wave quantization schemes are identical. Hence, these two
quantization schemes are unitary equivalent, see
Ref.~\cite{narozhny}.

The most remarkable property of boost modes is their behavior at the
light cone. It is easily seen from (\ref{scboostm}),
(\ref{scboostm1}) that at the vertex of the light cone
$\Psi_{\kappa}^{(\pm)}(x)$ possess the Dirac $\delta$-function
singularity
\begin{equation}\label{psi0}
\Psi_\kappa^{(\pm)}(0)=\frac{1}{\sqrt{2}}\,\delta(\kappa)\,.
\end{equation}
It was shown in Ref.~\cite{narozhny} that $\Psi_{\kappa}^{(\pm)}(x)$
possess a $\delta$-function singularity at the lines $x_{\pm}\equiv t\pm
z=0$ as well. Therefore contribution of the \emph{single} spectral
point $\kappa=0$ to physical quantities can be finite. We will
illustrate this by the example of Wightman function.

Taking into account the property of translational invariance we can
rewrite Eq.~(\ref{boost_compl}) in the form
\begin{equation}\label{trinv}
  \Delta^{(+)}(x'-x'')=i\int\limits_{-\infty}^\infty
  d\kappa\Psi_\kappa^{(+)}(x'){\Psi^{(+)}_\kappa}^*(x'')=
  i\int\limits_{-\infty}^\infty
  d\kappa\Psi_\kappa^{(+)}(x'-x''){\Psi^{(+)}_\kappa}^*(0)\,.
\end{equation}
Using now Eq.(\ref{psi0}) we obtain
\begin{equation}\label{psi01}
\Delta^{(+)}(x'-x'')=\frac{i}{\sqrt{2}}\,\Psi_0^{(+)}(x'-x'')\,,
\end{equation}
i.e. the integral over $\kappa$ in (\ref{trinv}) is determined by
the point $\kappa=0$ entirely\footnote{We have mentioned already
that Eq.~(\ref{psi01}) directly follows from Eqs.~(\ref{boostgen})
and (\ref{boostgen1}) accurate within a constant factor. Then
Eq.~(\ref{psi0}) could be derived from Eqs.~(\ref{psi01}),
(\ref{trinv}) even not using the representation (\ref{scboostm}).}.

This result means that the point $\kappa=0$ cannot be deleted from
the spectrum, or in other words the integral over $\kappa$ in
(\ref{boost_compl}),(\ref{trinv}) cannot be changed by its principal
value
\begin{equation}\label{pr}
\int\limits_{-\infty}^\infty d\kappa\ldots\neq {\rm
P.v.}\int\limits_{-\infty}^\infty
  d\kappa\ldots\,.
\end{equation}
Thereby, the family of boost modes does not constitute a complete
set in MS after excluding the zero mode. However, the equality in
Eq.~(\ref{pr}) could be restored if we cut the light cone out of MS,
compare \cite{narozhny}. This is because all the points where boost
modes possess $\delta(\kappa)$ singularity are located just at the
light cone.

Furthermore, the Wightman function Eq.~(\ref{trinv}) cannot be
represented in the form
\begin{equation}\label{trinv_err}
 \widetilde{\Delta}^{(+)}(x'-x'')=i\int\limits_0^\infty
  d\kappa\{\Psi_\kappa^{(+)}(x'){\Psi_\kappa^{(+)}}^*(x'')+
  \Psi_{-\kappa}^{(+)}(x'){\Psi_{-\kappa}^{(+)}(x'')}^*\}\,.
\end{equation}
If it were so, we could rewrite (\ref{trinv}), due to translational
invariance of the Wightman function and the boost mode property
(\ref{psi0}), as follows
\begin{equation}\label{trinv-err1}
 \widetilde{\Delta}^{(+)}(x'-x'')=\frac{i}{\sqrt{2}}\int\limits_0^\infty
  d\kappa\{\Psi_\kappa^{(+)}(x'-x'')+\Psi_{-\kappa}^{(+)}(x'-x'')\}
  \delta(\kappa)\,.
\end{equation}
But Eq.~(\ref{trinv-err1}) is evidently meaningless. Indeed, the
distribution $\delta(\kappa)$ is defined on the class of functions
continuous at the interval including the point $\kappa=0$. Therefore
Eq.~(\ref{trinv-err1}) should be understood as
\begin{equation}\label{trinv-err2}
 \widetilde{\Delta}^{(+)}(x'-x'')=\frac{i}{\sqrt{2}}\int\limits_{-\infty}^\infty
  d\kappa\{\Psi_\kappa^{(+)}(x'-x'')+\Psi_{-\kappa}^{(+)}(x'-x'')\}\theta(\kappa)
  \delta(\kappa)\,,
  \end{equation}
where $\theta(\kappa)$ is the Heaviside step function. However, the
product of two distributions $\theta(\kappa)\delta(\kappa)$ is not
defined.

The authors of Ref.~\cite{matsas} do not agree with this statement.
Discussing their Eq.~(2.129), which in our notations coincides with
(\ref{trinv_err}), they admit that this expression is undefined if
either of the two points $x',\,x''$ is located on the light cone.
They think that smearing of the distribution $\Delta^{(+)}(x'-x'')$
with compactly supported functions $f(x')$ and $g(x'')$ improve the
situation and the smeared "Wightman function" $\,$(2.131) is well
defined on the whole MS. Our analysis shows however that this is not
correct. Indeed, if we use Eq.~(\ref{trinv_err}) in the form
(\ref{trinv-err1}), we see that this expression is undefined for
\emph{arbitrary} $x',\,x''$. Moreover, smearing cannot improve the
situation since it does not influence the $\delta$-function in
(\ref{trinv-err1}). The correct expression for the smeared Wightman
function can be easily obtained from Eq.~(\ref{trinv}). It reads
\begin{equation}\label{sm_Wf}
\Delta^{(+)}(f,g)=\int d^2x'd^2x''f^*(x')g(x'')\Delta^{(+)}(x'-x'')=
i\int\limits_{-\infty}^\infty
   d\kappa\Psi_\kappa^{(+)}(f,g){\Psi_\kappa^{(+)}}^*(0)=
   \frac{i}{\sqrt{2}}\Psi_0^{(+)}(f,g)\,,
\end{equation}
where
\begin{equation}\label{sm_bm}
\Psi_\kappa^{(+)}(f,g)=\int d^2x'd^2x''f^*(x')g(x'')
\Psi_\kappa^{(+)}(x'-x'')\,,
\end{equation}
is the smeared boost mode. We see that again only the single
spectral point $\kappa=0$ contributes to the integral (\ref{sm_Wf})
for the smeared Wightman function.

Note, that we could try to use expression (\ref{trinv_err})
directly, not applying the property of translational invariance to
it. In that case we should attach exact mathematical meaning to
(\ref{trinv_err}) first. Namely, we should write it in the form
\begin{equation}\label{trinv_err_impr}
 \widetilde{\Delta}^{(+)}(x',x'')=i\lim\limits_{\varepsilon\to0}\int
 \limits_{\varepsilon}^\infty
  d\kappa\{\Psi_\kappa^{(+)}(x'){\Psi_\kappa^{(+)}}^*(x'')+
  \Psi_{-\kappa}^{(+)}(x'){\Psi_{-\kappa}^{(+)}}^*(x'')\}\,.
\end{equation}
But it can be easily seen that function
$\widetilde{\Delta}^{(+)}(x',x'')$ in (\ref{trinv_err_impr}) is not
translationally invariant and thus has nothing to do with Wightman
function for a free field in MS. Indeed, if it were translationally
invariant, it could be written down as
\begin{equation}\label{trinv_err_impr1}
\widetilde{\Delta}^{(+)}(x',x'')=\frac{i}{\sqrt{2}}\lim\limits_{\varepsilon\to0}\int
 \limits_{\varepsilon}^\infty
  d\kappa\{\Psi_\kappa^{(+)}(x'-x'')+\Psi_{-\kappa}^{(+)}(x'-x'')\}\delta(\kappa)\,,
\end{equation}
and would be equal to zero identically.

Singular behavior of the boost mode at the point $\kappa=0$ can be
understood in terms of classical trajectories of free particles with
a given value of the boost parameter $\kappa$
\begin{equation} \label{kap}
  \kappa=z\varepsilon_p-tp,
\end{equation}
where $p=m\dot{z}/\sqrt{1-\dot{z}^2}$ is the momentum,
$\varepsilon_p=m/\sqrt{1-\dot{z}^2}$ -- the energy of a particle,
$\dot{z}=dz/dt$. Let us rewrite Eq.~(\ref{kap}) in the the form
\begin{equation}\label{traj}
  z=\dot{z}t+\frac{\kappa}{m}\sqrt{1-\dot{z}^2}.
\end{equation}
\begin{figure}
\includegraphics[scale=0.7]{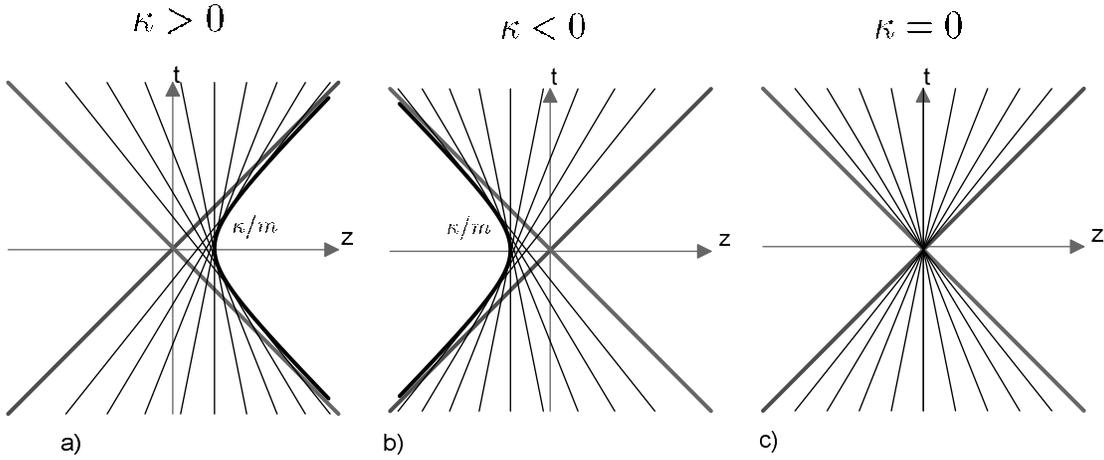}
\caption{\small Classical trajectories of massive boost particles.}
\end{figure}
After differentiating both sides of Eq.~(\ref{traj}) with time one
can easily find its regular
\begin{equation}\label{reg}
  z=vt+\frac{\kappa}{m}\sqrt{1-v^2},\qquad v=const\,,\quad |v|<1\,,
\end{equation}
and singular
\begin{equation}\label{spec}
  z=\mathrm{sgn}(\kappa)\sqrt{t^2+\frac{\kappa^2}{m^2}}\,,
\end{equation}
solutions.

When $\kappa\neq0$ the singular solution is represented by two
branches of hyperbola (\ref{spec}). The branch with $\kappa>0$ is
located in the right wedge of MS, while the branch with $\kappa<0$
in the left one, see Figs.~1a,b. Free boost particles with the given
value of $\kappa$ propagate along the regular world lines which are
different straight lines tangent to the corresponding branches of
hyperbola (\ref{spec}). It is worth noting that classical particles
with positive values of the boost parameter $\kappa$ cannot
penetrate to the left wedge, as well as particles with $\kappa<0$
cannot penetrate to the right one.

The world lines of boost particles with $\kappa=0$ are given by
\begin{equation}\label{trajk0}
  z=vt\,.
\end{equation}
All of them cross the vertex of the light cone $h_0$
and any world line crossing $h_0$ belongs to the family of
trajectories with $\kappa=0$ , see Fig.~1c. Hence, elimination of the point
$\kappa=0$ from the spectrum is equivalent to prohibition for
particles to cross the point $h_0$, or to pricking it out of MS.
Besides, deleting the point $\kappa=0$ from the spectrum, we delete
a bunch of \emph{infinite} number of trajectories, thus loosing a
substantial part of degrees of freedom. This means that the point
$\kappa=0$ is of nonzero measure at $h_0$ and explains the
$\delta$-function singularity of the boost mode there.

The points lying at the light cone surface need a separate
consideration. The lines $z=\pm t$ cannot be trajectories of massive
particles. But the branches of hyperbola (\ref{spec}) degenerate
into lines constituting the surface of the light cone in our 2d
problem when $\kappa\rightarrow0$. As it can easily be seen from
Figs.~1a,b, all regular trajectories except those which are tangent
to the hyperbola branches at the point $t=0$ also tend to the lines
$z=\pm t$ in the limit $\kappa\rightarrow0$. Hence, these lines
$z=\pm t$ are the lines of condensation for regular trajectories
with $\kappa\rightarrow0$. Possibly this explains the existence of
$\delta$-function singularity of boost modes at the light cone
surface.

It is worth noting that there exist another representation for the
boost modes (\ref{scboostm}), (\ref{scboostm1}). Keeping in mind
that an infinitely small negative imaginary part $-i\sigma$ is added
to $t$ in (\ref{scboostm}), after the change of the variable of
integration $q=\ln u$ we obtain for $\Psi_\kappa^{(\pm)}(x)$ the
following expression
\begin{equation}
\label{scbm2}\displaystyle
\Psi_\kappa^{(\pm)}(x)=\frac{1}{2^{3/2}\pi} \int\limits_{0}^\infty
du u^{-i\kappa-1} e^{-\frac{\beta}{u}-\gamma u} \,,
\end{equation}
with $\displaystyle\beta=\frac{m}{2}(\sigma\pm ix_+)$ and
$\gamma=\frac{m}{2}(\sigma\pm ix_-)$. Using now formula 3.471(9) of
Ref.~\cite{GrR} we get
\begin{equation} \label{scbm3}
\Psi_\kappa^{(\pm)}(x)=\frac{1}{\pi\sqrt{2}}
\displaystyle\left(\frac{x_-\mp i\sigma}{x_+\mp i\sigma}
\right)^{i\frac{\kappa}{2}} K_{i\kappa}(w_{\pm})\,,\qquad
w_{\pm}=m\sqrt{e^{\pm i\pi}(x_+\mp i\sigma)(x_-\mp i\sigma)}\,,
\end{equation}
where $K_{i\kappa}(w_{\pm})$ is the Macdonald function, and
distributions $(\xi\mp i\sigma)^{\lambda}$ should be understood as
\begin{equation}
\label{i0}
    (\xi\mp i\sigma)^{\lambda}=\xi^{\lambda}\theta(\xi)+e^{\mp i\pi
    \lambda}(-\xi)^{\lambda}\theta(-\xi),
\end{equation}
see Ref.~\cite{gelfand}. Using Eq.~(\ref{i0}) we obtain
\begin{equation}\label{scbmsec}
  \Psi_\kappa(x)=\theta(-x_+)\theta(-x_-)\Psi^{(P)}_\kappa(x)+
  \theta(x_+)\theta(-x_-)\Psi^{(R)}_\kappa(x)+
  \theta(x_+)\theta(x_-)\Psi^{(F)}_\kappa(x)+\theta(-x_+)
  \theta(x_-)\Psi^{(L)}_\kappa(x),
\end{equation}
see \cite{gerlach,narozhny}, where
\begin{equation}\label{scbmsec1}
\begin{split}
  &\Psi^{(P)}_\kappa(x)=\frac{ie^{-\pi\kappa/2}}{2^{3/2}}
  \left(\frac{-x_-}{-x_+}\right)^{i\kappa/2}H^{(1)}_{i\kappa}
  \left(m\sqrt{(-x_+)(-x_-)}\right),\quad
  \Psi^{(R)}_\kappa(x)=\frac{e^{\pi\kappa/2}}{\pi\sqrt{2}}
  \left(\frac{-x_-}{x_+}\right)^{i\kappa/2}K_{i\kappa}
  \left(m\sqrt{x_+(-x_-)}\right)\\
  &\Psi^{(F)}_\kappa(x)=\frac{-ie^{\pi\kappa/2}}{2^{3/2}}
  \left(\frac{x_-}{x_+}\right)^{i\kappa/2}H^{(2)}_{i\kappa}
  \left(m\sqrt{x_+x_-}\right),\quad
  \Psi^{(L)}_\kappa(x)=\frac{e^{-\pi\kappa/2}}{\pi\sqrt{2}}
  \left(\frac{x_-}{-x_+}\right)^{i\kappa/2}K_{i\kappa}
  \left(m\sqrt{(-x_+)x_-}\right)
\end{split}
\end{equation}
are the expressions for the boost modes in the past ($P$), right
($R$), future ($F$) and left ($L$) wedges of MS respectively. Here
$H^{(1,2)}_{i\kappa}(w)$ are Hankel functions.

Putting $\kappa=0$ in Eq.~(\ref{scbmsec1}) and taking into account
Eq.~(\ref{psi01}) one can easily reproduce the Wightman result
\cite{wightman} for $\Delta^{(+)}(x)$. Applying the same procedure
to Eq.~(\ref{scbm3}) we get
\begin{equation}\label{Wf_s}
\Delta^{(\pm)}(x)=\pm \frac{i}{2\pi}K_0(w_{\pm})\,.
\end{equation}
This is a new compact representation for the Wightman function in 2d
scalar theory.

\section{Fermions}

A set of orthonormalized positive (negative) frequency solutions
$\{F_a^{(\pm)}(x)\}$ of Dirac equation in 2d MS
\begin{equation}\label{direq}
\displaystyle \mathcal{D}_-\psi(x)=0\,,\,\,\quad \mathcal{D}_{\pm}=
  \left(i\gamma^0\frac{\partial}{\partial t}+i\gamma^1\frac{\partial}{\partial
  z}\pm m\right)\,,
\end{equation}
($\gamma^{0,1}$ -- two-dimensional Dirac matrices,
$\gamma^0\equiv\beta$,
$\gamma^0\gamma^1=-\gamma^1\gamma^0\equiv\alpha$,
$\alpha^2=\beta^2=1$) is complete if
\begin{equation}\label{fpolnota}
  \int\limits da F_a^{(\pm)}(x'){F_a^{(\pm)}}^{\dag}(x'')\gamma^0=-iS^{(\pm)}(x'-x''),
\end{equation}
where the positive $(+)$ and negative $(-)$ frequency Wightman
functions for the fermion field are equal to
\begin{equation}\label{wfscferm}
  S^{(\pm)}(x)=\mathcal{D}_+\Delta^{(\pm)}(x),\quad
  \Delta^{(-)}(x)=\Delta^{(+)^*}(x)\,,
\end{equation}
and $\Delta^{(+)}(x)$ is defined in (\ref{scwf1}).

Any set $\{F_a^{(\pm)}(x)\}$ can be constructed of the set of
matrices $\{S^{(\pm)}(x-u)\}$ in the perfect analogy with the boson
case. Taking into account that the Dirac operators $\mathcal{D}_\pm$
commute with the fermion boost operator $\mathcal{B_D}$,
\begin{equation}\label{genbf}
[\mathcal{D}_\pm,\mathcal{B_D}]_-=0,\quad \mathcal{B_D}=
\mathcal{B_K}-\frac{i}{2}\gamma^0\gamma^1=i\left(z\partial_t+
t\partial_z-\frac{1}{2}\alpha\right)\,,
\end{equation}
it is easy to check that functions
\begin{equation}\label{fermrazl}
\psi^{(\pm)}_\kappa(x)=\int\limits_{-\infty}^\infty dq S^{(\pm)}(x-u_q)f_\kappa(q)
\end{equation}
($u_q$ are defined in (\ref{alph})) are the eigenfunctions of this
operator if
\begin{equation}\label{f_k}
f_\kappa(q)=e^{-i\kappa q+\frac{1}{2}\alpha q}C,
\end{equation}
where  $C$ is an arbitrary constant column. Substituting (\ref{f_k})
into Eq.~(\ref{fermrazl}) we easily get with the account of
definition (\ref{wfscferm}) the following expression for the
normalized boost modes
\begin{equation}\label{fermbmgen}
\psi_\kappa^{(\pm)}(x)=\frac{1}{2\pi}\sqrt{\frac{m}{2}}\int
\limits_{-\infty}^\infty dq e^{\mp im[(t\mp i\sigma)\cosh q-z\sinh q]-
i\kappa q+\frac{1}{2}\alpha q}(1\pm\beta)\eta.
\end{equation}
Here the arbitrary constant column $\eta$ satisfies the relation
\begin{equation}
\eta^+(1\pm\beta)\eta=1.
\end{equation}

It is difficult to compare representation (\ref{fermbmgen}) with the
results of Refs.\cite{muller,mullerbook,McMah} directly since the
authors of these works by unknown reasons use in 2d spacetime
$4\times4$ matrices which constitute a reducible representation of
the Dirac matrices. The passage to their representation may be
realized as follows
\begin{equation}\label{2_4}
\alpha\rightarrow\alpha_3=\alpha\otimes\sigma_3,\quad
\beta\rightarrow\beta\otimes I,\quad
\eta\rightarrow\eta\otimes\begin{pmatrix}1\\0\end{pmatrix},
\end{equation}
where
\begin{equation}\label{alpha1}
\alpha=\begin{pmatrix}&0&1\\&1&0\end{pmatrix},\quad \beta=
\begin{pmatrix}&1&0\\&0&-1\end{pmatrix},\quad \eta=\frac{1}{\sqrt{2}}
\begin{pmatrix}1\\1\end{pmatrix}.
\end{equation}
In particular, using the notations of Refs.~\cite{muller,mullerbook}
\begin{equation}\label{mobozn}
v=\frac{1}{2}\ln\frac{x_+}{-x_-},\quad u=\sqrt{x_+(-x_-)},\quad
\kappa\equiv\omega,\quad
\Phi^\pm_\omega=H_{i\omega\pm1/2}^{(1)}(imu),
\end{equation}
for the positive frequency solution in the Rindler wedge of 2d MS we
obtain from Eq.~(\ref{fermbmgen}) with account of
(\ref{2_4}),(\ref{alpha1})
\begin{equation}\label{fbmmul}
\psi_\omega^+\theta(x_+)\theta(-x_-)=\frac{i}{4}\sqrt{m}e^{-i\omega
v}
\begin{pmatrix}e^{\frac{1}{2}v}\Phi^-_\omega+e^{-\frac{1}{2}v}
\Phi^+_\omega\\0\\e^{\frac{1}{2}v}\Phi^-_\omega-e^{-\frac{1}{2}v}\Phi^+_
\omega\\0\end{pmatrix}\theta(x_+)\theta(-x_-).
\end{equation}

It is worth noting that the results of
Refs.~\cite{muller,mullerbook,McMah}, see Eqs.  21.64, 21.65, 21.77
and 21.78à in \cite{mullerbook}, differ from (\ref{fbmmul}) by the
absence of factors $e^{\pm\frac{1}{2}v}$ in front of the functions
$\Phi_\omega^\mp$ in (\ref{fbmmul}). In the framework of the method
used by the authors of Refs.~\cite{muller,mullerbook}, the latter
should have appeared as a result of change of Rindler $\{u,v\}$ by
Cartesian coordinates
\begin{equation}\label{dovorot}
 \psi(t,z)=e^{\frac{1}{2}\alpha v}\psi(u,v),
\end{equation}
see, e.g., \cite{fock}.

Besides, the modes in Refs.~\cite{muller,mullerbook,McMah} differ
from (\ref{fbmmul}) by additional normalization factor $1/\sqrt{3}$.
This discrepancy has arisen due to incorrect method of normalization
used by the authors of Refs.~\cite{muller,mullerbook,McMah}. Indeed,
they have represented the normalization integral in 2d MS as a sum
of four terms each equal to the normalization integral in one of
four wedges of MS. However, the correct normalization procedure
implies integration over some Cauchy surface in MS. The most
convenient surface for this purpose is the surface $t=0$, compare,
e.g., \cite{unruh,narozhny}. Then only the solutions in $R$ and $L$
wedges of MS contribute to the normalization integral and the
normalization factor coincides with that one in (\ref{fbmmul}).

From now on, we will use a representation of Dirac matrices
different from (\ref{alpha1}),
\begin{equation}\label{diracm}
\gamma^0\equiv\beta=\sigma_1,\gamma^1=i\sigma_2,\gamma^0\gamma^1
\equiv\alpha=-\sigma_3,
\end{equation}
where $\sigma_i$, $i=1,2,3$ are Pauli matrices.  Use of
representation (\ref{diracm}) allows  essentially simplify the form
of solutions of Eq.~(\ref{direq}). We will see below that it also is
very convenient for the procedure of transition to the limit
$m\rightarrow0$. With this set of Dirac matrices the representation
(\ref{fermbmgen}) for the boost modes reduces to
\begin{equation}
\label{fermbm}\displaystyle
\psi_\kappa^{(\pm)}(x)=\frac{1}{2\pi}\sqrt{\frac{m}{2}}
\int\limits_{-\infty}^\infty dq e^{\mp im((t\mp i\sigma)\cosh q-z\sinh q)-
i\kappa q}
\begin{pmatrix}
e^{-q/2}\\
\pm e^{q/2}
\end{pmatrix}.
\end{equation}
Using exactly the same procedure as for the scalar boost modes
(\ref{scboostm}) we can transform Eq.~(\ref{fermbm}) to the form
\begin{equation}
\label{fermbm2}
\psi_\kappa^{(\pm)}(x)=\frac{1}{\pi}\sqrt{\frac{m}{2}}
\begin{pmatrix}
\displaystyle\left(\frac{x_-\mp i\sigma}{x_+\mp i\sigma}
\right)^{\frac{1}{4}+i\frac{\kappa}{2}}
K_{\frac{1}{2}+i\kappa}(w_{\pm})\\
\displaystyle\pm \left(\frac{x_-\mp i\sigma}{x_+\mp i\sigma}
\right)^{-\frac{1}{4}+i\frac{\kappa}{2}}
K_{\frac{1}{2}-i\kappa}(w_{\pm})
\end{pmatrix}\,,
\end{equation}
where again $w_{\pm}=m\sqrt{e^{\pm i\pi}(x_+\mp i\sigma)(x_-\mp
i\sigma)}\,$.

In perfect analogy with Eq.~(\ref{scbmsec}) the positive frequency
boost mode (\ref{fermbm2}) can be rewritten in the form
\begin{equation}\label{fr}
  \psi^{(+)}_\kappa(x)=\theta(-x_+)\theta(-x_-){}^P\!\psi^{(+)}_\kappa(x)+
  \theta(x_+)\theta(-x_-){}^R\!\psi^{(+)}_\kappa(x)+
  \theta(x_+)\theta(x_-){}^F\!\psi^{(+)}_\kappa(x)+\theta(-x_+)
  \theta(x_-){}^L\!\psi^{(+)}_\kappa(x),
\end{equation}
where in $P$ wedge
\begin{equation}\label{fermbmp}
\displaystyle
  {}^P\!\psi^{(+)}_\kappa(x)=\frac{i}{2}\sqrt{\frac{m}{2}}e^
  {-\frac{\pi\kappa}{2}+\frac{i\pi}{4}}
  \begin{pmatrix}\displaystyle
  \left(\frac{-x_-}{-x_+}\right)^{\frac{1}{4}+\frac{i}{2}\kappa}
  H^{(1)}_{1/2+i\kappa}\left(m\sqrt{(-x_+)(-x_-)}\right)\\ \displaystyle
  -i\left(\frac{-x_-}{-x_+}\right)^{-\frac{1}{4}+\frac{i}{2}\kappa}
  H^{(1)}_{-1/2+i\kappa}\left(m\sqrt{(-x_+)(-x_-)}\right)
  \end{pmatrix},
\end{equation}
in $R$ wedge
\begin{equation}\label{fermbmr}
\displaystyle
  {}^R\!\psi^{(+)}_\kappa(x)=\frac{1}{\pi}\sqrt{\frac{m}{2}}e^
  {\frac{\pi\kappa}{2}-\frac{i\pi}{4}}
  \begin{pmatrix}\displaystyle
  \left(\frac{-x_-}{x_+}\right)^{\frac{1}{4}+\frac{i}{2}\kappa}
  K_{1/2+i\kappa}\left(m\sqrt{x_+(-x_-)}\right)\\ \displaystyle
  i  \left(\frac{-x_-}{x_+}\right)^{-\frac{1}{4}+\frac{i}{2}\kappa}
  K_{-1/2+i\kappa}\left(m\sqrt{x_+(-x_-)}\right)
  \end{pmatrix},
\end{equation}
in $F$ wedge
\begin{equation}\label{fermbmf}
\displaystyle
  {}^F\!\psi^{(+)}_\kappa(x)=-\frac{i}{2}\sqrt{\frac{m}{2}}e^
  {\frac{\pi\kappa}{2}-\frac{i\pi}{4}}
  \begin{pmatrix}\displaystyle \left(\frac{x_-}{x_+}\right)^
  {\frac{1}{4}+\frac{i}{2}\kappa} H^{(2)}_{1/2+i\kappa}
  \left(m\sqrt{x_+x_-}\right)\\ \displaystyle
  i\left(\frac{x_-}{x_+}\right)^{-\frac{1}{4}+\frac{i}{2}\kappa}
  H^{(2)}_{-1/2+i\kappa}\left(m\sqrt{x_+x_-}\right)
  \end{pmatrix},
\end{equation}
and finally in $L$ wedge
\begin{equation}\label{fermbml}
\displaystyle
  {}^L\!\psi^{(+)}_\kappa(x)=\frac{1}{\pi}\sqrt{\frac{m}{2}}
  e^{-\frac{\pi\kappa}{2}+\frac{i\pi}{4}}
  \begin{pmatrix}\displaystyle \left(\frac{x_-}{-x_+}\right)^{\frac{1}{4}+\frac{i}{2}
  \kappa} K_{1/2+i\kappa}\left(m\sqrt{(-x_+)x_-}\right)\\ \displaystyle
  -i\left(\frac{x_-}{-x_+}\right)^{-\frac{1}{4}+\frac{i}{2}\kappa}
  K_{-1/2+i\kappa}\left(m\sqrt{(-x_+)x_-}\right)
  \end{pmatrix}.
\end{equation}
A similar representation is valid for the negative frequency modes
$\psi_\kappa^{(-)}(x)$ as well.

The boost modes (\ref{fermbm}), (\ref{fermbm2}) are orthonormalized
by the condition
\begin{equation}\label{normf}
  (\psi_{\kappa'}^{(\pm)},\psi_{\kappa}^{(\pm)})=
  \int\limits_{-\infty}^\infty dz{\psi_{\kappa'}^{(\pm)}}^\dag(x)
  \psi_{{\kappa}}^{(\pm)}(x)=\delta(\kappa-\kappa'),
\end{equation}
constitute a complete set in the sense (\ref{fpolnota}), and thus
can serve a basis for quantization of the fermion field in MS,
\begin{equation}\label{fermkvant}
  \Psi(x)=\int\limits_{-\infty}^\infty
  d\kappa\left(b_\kappa\psi_\kappa^{(+)}(x)+\tilde
  b^+_\kappa\psi_\kappa^{(-)}(x)\right),
\end{equation}
where $b_\kappa$ and $\tilde b_\kappa$ are annihilation operators
for boost fermion particles and antiparticles respectively, which
obey the standard commutation relations. By the same reasons as in
the boson case, boost quantization of the fermion field is unitary
equivalent to quantization in the plane wave basis
\begin{equation}\label{fermplanekv}
 \Psi(x)=\int\limits_{-\infty}^\infty
  dp\left(a_p\psi_p^{(+)}(x)+\tilde
 a_p^+\psi_p^{(-)}(x)\right),
\end{equation}
where
\begin{equation}\label{psipsipferm}
  \psi_p^{(\pm)}(x)=\begin{pmatrix}\sqrt{\varepsilon_p-
  p}\\\pm\sqrt{\varepsilon_p+p}\end{pmatrix}\frac{e^{\mp
  i\varepsilon_pt\pm ipz}}{\sqrt{4\pi\varepsilon_p}},\quad\quad
  \varepsilon_p=\sqrt{p^2+m^2}.
\end{equation}
It is not difficult to ascertain that the boost modes (\ref{fermbm})
are connected with the plane waves (\ref{psipsipferm}) through the
following integral transformation
\begin{equation}\label{boostplane}
  \psi_\kappa^{(\pm)}(x)=\int\limits_{-\infty}^\infty
  \frac{dp}{\sqrt{2\pi\varepsilon_p}}e^{- iq\kappa}\psi^{(\pm)}_p(x)
 ,\quad\quad q=\mathrm{arcsinh}\,\frac{p}{m}.
\end{equation}

The modes $\psi^{(\pm)}_p(x)$ and $\psi^{(\pm)}_\kappa(x)$ are
distributions with respect to variables $p$ and $\kappa$. This means
that they define linear functionals
\begin{equation}\label{fp}
  \mathfrak{F}^{(\pm)}(\mathfrak{g},x)=\int\limits_{-\infty}^\infty
  dp\,\psi^{(\pm)}_p(x)\mathfrak{g}(p),
\end{equation}
\begin{equation}\label{fk}
 \mathfrak{ F}^{(\pm)}(\mathfrak{h},x)=\int\limits_{-\infty}^\infty
  d\kappa\,\psi^{(\pm)}_\kappa(x)\mathfrak{h}(\kappa),
\end{equation}
on some sets of test functions $\mathfrak{g}(p)$,
$\mathfrak{h}(\kappa)$ respectively. Such functionals naturally
appear when we calculate, e.g., matrix elements of the field
operators (\ref{fermkvant}), (\ref{fermplanekv}) between the vacuum
state $|0_M\rangle$ and the states which are particle (or
antiparticle) wave packets,
\begin{equation}\label{1f}
  |1_\mathfrak{g}\rangle=\int\limits_{-\infty}^\infty dp\,\mathfrak{g}(p)a^+_p
  |0_M\rangle, \quad |1_\mathfrak{h}\rangle=\int\limits_{-\infty}^\infty
  d\kappa\,\mathfrak{h}(\kappa)b^+_\kappa
  |0_M\rangle.
\end{equation}

Physical one-particle states must be normalized,
\begin{equation}\label{1_norma}
  \langle 1_\mathfrak{g}|1_\mathfrak{g}\rangle=\int\limits_{-\infty}^{\infty}dp|\mathfrak{g}(p)|^2
  =1,\quad\langle 1_\mathfrak{h}|1_\mathfrak{h}\rangle=\int\limits_{-\infty}^
  {\infty}d\kappa|\mathfrak{h}(\kappa)|^2=1.
\end{equation}
Hence, $\mathfrak{g}(p)$, $\mathfrak{h}(\kappa)$  must be
square-integrable functions of $p$ and $\kappa$ respectively. Then
it is easy to see that matrix elements (\ref{fp}), (\ref{fk}) are
square-integrable functions of $x$. It is clear that functionals
(\ref{fp}), (\ref{fk}) are the elements of one and the same
space of functions. Hence, there exists one to one correspondence
between the test functions $\mathfrak{g}(p)$ and $\mathfrak{h}(\kappa)$:
\begin{equation}
\label{1} \mathfrak{h}(\kappa)=\int\limits_{-\infty}^\infty \frac{e^{i\kappa
q}}{\sqrt{2\pi\varepsilon_p}}\mathfrak{g}(p) dp\,,
\end{equation}
where $q=\mathrm{arsinh}(p/m)$ is rapidity. For the sake of
convenience we will confine ourselves to $\mathfrak{g}(p)$ belonging
to the class of continuous piecewise smooth functions descending
faster than $|p\,|^{-1}$ at $|p\,|\rightarrow\infty$. This
requirement guarantees finiteness of the mean value of the energy of
the one-particle state $\left|1_\mathfrak{g}\right>$ (\ref{1f}),
\begin{equation}
\label{e1}
  E(1)=\int\limits_{-\infty}^\infty dp\,\varepsilon_p
  \,\left<1_\mathfrak{g}\right|a_p^+ a_p \left|1_\mathfrak{g}\right>=\int\limits_{-\infty}^\infty
  dp\,\varepsilon_p\,|\mathfrak{g}(p)|^2,\quad \varepsilon_p=\sqrt{p^2+m^2},
\end{equation}
To ascertain the properties of test functions $\mathfrak{h}(\kappa)$
we will use the relation (\ref{1}).

First, we will illustrate by a simple example that, unlike the case
of the functional (\ref{fp}), we need know the behavior of test
functions $\mathfrak{h}(\kappa)$ for the functional (\ref{fk}) not
only on the real axis but also in the complex plane. Consider
\begin{equation}\label{gaussp}\displaystyle
  \mathfrak{g}(p)=\frac{e^{-\frac{\beta^2}{4\alpha}}}{(2\pi\alpha{\varepsilon_p}^2)^{1/4}}\,
 e^{-\frac{q^2}{4\alpha}+\frac{\beta
  q}{2\alpha}},\quad\quad\alpha>0\,,
\end{equation}
which evidently satisfies the above formulated requirements to
functions $\mathfrak{g}(p)$. Using Eq.~(\ref{1}) we obtain
\begin{equation}\label{hgauss}
 \mathfrak{h}(\kappa)=\left(\frac{2\alpha}{\pi}\right)^{1/4}e^{-\alpha\kappa^2+ i\beta\kappa}.
  \end{equation}
Direct calculation of the integral (\ref{fk}) at the point $x=0$
then yields
\begin{equation}\label{fkgauss}
 \mathfrak{F}^{(\pm)}(\mathfrak{h},0)=\sqrt{\frac{m}{2}}\left(\frac{2\alpha}{\pi}\right)^{1/4}
  \begin{pmatrix} e^{  \frac{\alpha}{4}-\frac{\beta}{2}}\\
  \displaystyle\pm e^{\frac{\alpha}{4}+\frac{\beta}{2}}\end{pmatrix}=
  \sqrt{\frac{m}{2}}\begin{pmatrix}\mathfrak{h}(i/2)\\\pm
  \mathfrak{h}(-i/2)\end{pmatrix}.
\end{equation}
So we see that the value of the functional (\ref{fk}) at the vertex
of the light cone is determined by the values of the test function
$\mathfrak{h}(\kappa)$ at imaginary points $\kappa=\pm i/2$.

Let us now revert to Eq.~(\ref{1}). After the change of the variable
of integration $p=m\sinh q$ it can be written as a Fourier transform
of function
$\widetilde{\mathfrak{g}}(q)=\displaystyle\sqrt{\frac{m\cosh
q}{2\pi}}\,\mathfrak{g}(m\sinh q)$
\begin{equation}\label{F_tr}
\mathfrak{h}(\kappa)=\int\limits_{-\infty}^\infty dq e^{i\kappa q }
\,\widetilde{\mathfrak{g}}(q)\,.
\end{equation}
Under our assumptions
\begin{equation}\label{f_as}
\widetilde{\mathfrak{g}}(q)=O\left(e^{-(\frac{1}{2}+\epsilon)|q|}\right)
\,,\,\, \epsilon>0,
\end{equation}
as $|q|\rightarrow \infty\,$. Then, according to the Paley-Wiener
theorem \cite{W-P} the test functions $\mathfrak{h}(\kappa)$
(\ref{F_tr}) are analytic in the strip
\begin{equation}\label{polosa}
  -1/2-\epsilon<\mathrm{Im}\, \kappa<1/2+\epsilon, \quad\epsilon>0.
\end{equation}

Let us calculate the integral in Eq.~(\ref{F_tr}) by parts. Since
the test functions $\widetilde{\mathfrak{g}}(q)$ are continuous and
descend exponentially as $|q|\rightarrow\infty$, see
Eq.~(\ref{f_as}), the first integrated term is equal to zero. Hence,
the functions $\mathfrak{h}(\kappa)$ in their domain of analyticity
(\ref{polosa}) descend at least as $\kappa^{-2}$ when
$|\kappa|\rightarrow\infty\,$. This means that the path of
integration in (\ref{fk}) can be shifted in the range of the strip
(\ref{polosa}).

Consider the functional (\ref{fk}) at the point $x=0$. Shifting the
path of integration for the upper component upward, and for the
lower one downward, by $i/2$, we obtain
\begin{equation}\label{defdel}
  \mathfrak{F}^{(\pm)}(\mathfrak{h},0)=
  \frac{1}{2\pi}\sqrt{\frac{m}{2}}\int\limits_{-\infty}^\infty d\kappa
  dq e^{-i\kappa q}\begin{pmatrix}\mathfrak{h}(\kappa+i/2)\\\pm \mathfrak{h}(\kappa-i/2)
  \end{pmatrix}=
  \sqrt{\frac{m}{2}}\int\limits_{-\infty}^\infty d\kappa
  \delta(\kappa)\begin{pmatrix}\mathfrak{h}(\kappa+i/2)\\\pm \mathfrak{h}(\kappa-i/2)
  \end{pmatrix}=
  \sqrt{\frac{m}{2}}\begin{pmatrix}\mathfrak{h}(
  i/2)\\\pm \mathfrak{h}( -i/2)\end{pmatrix}.
\end{equation}
Eq.~(\ref{defdel}) generalizes formula (\ref{fkgauss}) to the class
of functions $\mathfrak{h}(\kappa)$ analytic in the strip
(\ref{polosa}) and descending rapidly enough when
$|\kappa|\to\infty$.

The obtained result makes it possible to write down the fermion
boost mode in the vertex of the light cone in terms of
$\delta$-functions of complex argument
\begin{equation}\label{fermdelta}
  \psi_\kappa^{(\pm)}(0)=\sqrt{\frac{m}{2}}\begin{pmatrix}\delta(\kappa-
  i/2)\\ \pm\delta(\kappa+i/2)\end{pmatrix}.
\end{equation}

\begin{figure}[t]
\epsfxsize14cm\epsffile{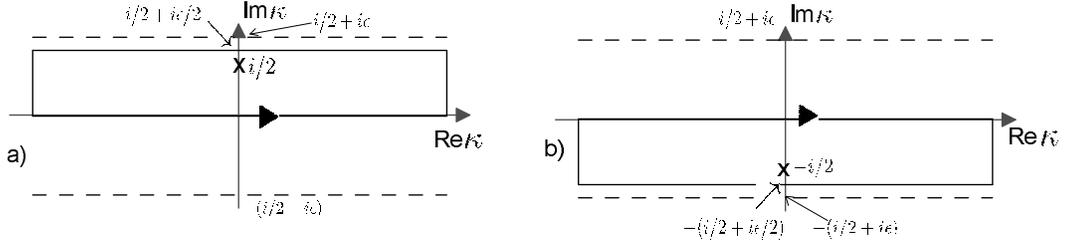}
\caption{\small Contours of integration for the upper (a) and lower (b)
components of $\mathfrak{F}^{(\pm)}(\mathfrak{h},0)$.}
\label{kontur_1}
\end{figure}

The functionals $\mathfrak{F}^{(\pm)}(\mathfrak{h},0)$ can be also
written as integrals around the closed contours represented in
Figs.~\ref{kontur_1}a,b
\begin{equation}\label{kont_1}
\mathfrak{F}^{(\pm)}(\mathfrak{h},0)=\oint\limits_{C_{a,b}}d\kappa
\psi_\kappa^{(\pm)}(0)\mathfrak{h}(\kappa)\,.
\end{equation}
Indeed, the explicit form for $\psi_\kappa^{(\pm)}(0)$ can be easily
derived from Eq.~(\ref{fermbm2}). Putting $x_+=x_-=0$ there and
using the ascending series for $K_{\nu}(z)$ \cite{GrR}, we get
\begin{equation}\displaystyle\label{psi0gamma}
 \psi_\kappa^{(\pm)}(0)=\lim\limits_{\sigma\to0}\frac{1}{2\pi}
 \begin{pmatrix}
\displaystyle
  \left(\frac{m}{2}\right)^{-i\kappa}\frac{\Gamma(1/2+ i\kappa)}
  {\sigma^{1/2+ i\kappa}}\\ \displaystyle
  \pm\left(\frac{m}{2}\right)^{i\kappa}\frac{\Gamma(1/2- i\kappa)}
  {\sigma^{1/2- i\kappa}}
 \end{pmatrix}.
\end{equation}
Then, taking into account that the integrand in (\ref{kont_1}) has
simple poles at $\kappa=\pm i/2$ and using the Cauchy's residue
theorem, we immediately reproduce the result (\ref{defdel}).
 Since the integrals taken along the
vertical segments of the contours $C_{a,b}$ are evidently equal to
zero, this means, in particular, that the integrals along the upper
(lower) piece of the contour $C_{a}$ ($C_{b}$) is also equal to zero
in the limit $\sigma\rightarrow 0$. The latter statement can be
easily checked by direct calculation.

$\delta$-function of complex argument was first introduced by
Gel'fand and Shilov in Ref.~\cite{gelfand} on the basis of Cauchy
residue theorem,
\begin{equation}
\label{ddef}
    (\delta(\kappa-\kappa_0),f(\kappa))=\frac{1}{2\pi
    i}\oint\limits_L\frac{f(\kappa)d\kappa}{\kappa-\kappa_0}=f(\kappa_0),
\end{equation}
where $L$ is a contour enclosing an arbitrary complex point $\kappa_0$. The
distribution (\ref{ddef}) was defined in \cite{gelfand} on some
class of entire functions $f(\kappa)$, the so called $Z$-class. Our
class of test functions analytic in the strip (\ref{polosa}) differs
from $Z$. Therefore some properties of the Gel'fand's
$\delta$-function \cite{gelfand} cannot be applied to our one.
However, the opposite statement is true: all properties of our
$\delta$-function hold valid for the Gel'fand's one.

So, on the class of test functions analytic in the strip
(\ref{polosa}), as well as for the $Z$-class, the following equality
of distributions is valid \cite{jetpl}
\begin{equation}\label{delta}
\displaystyle  \lim\limits_{\sigma\to0}\frac{1}{2\pi}
\frac{\Gamma(1/2\pm i\kappa)}
  {\sigma^{1/2\pm i\kappa}}=\delta\left(\kappa\mp i/2\right)\,.
\end{equation}

Let us show now that the modes (\ref{fermbm}), (\ref{fermbm2})
possess $\delta$-function singularities not only at the vertex of
the light cone but at lines $x_{\pm}=0$ as well. Indeed, at the
surface of the light cone the arguments of Macdonald functions in
(\ref{fermbm2}) are small, $|w_{\pm}|\ll 1$. Therefore using the
ascending series for Macdonald functions \cite{GrR} we get
\begin{equation}
\displaystyle \label{psic}
 \psi_\kappa^{(\pm)}(x)=\frac{e^{\pm\pi\kappa/2\mp i\pi/4}}{2\pi}
 \begin{pmatrix}
\displaystyle
  \left(\frac{m}{2}\right)^{-i\kappa}\frac{\Gamma(1/2+ i\kappa)}
  {(x_+\mp i\sigma)^{1/2+ i\kappa}}+
  \left(\frac{m}{2}\right)^{i\kappa+1}\frac{\Gamma(-1/2- i\kappa)}
  {(-x_-\pm i\sigma)^{-1/2-i\kappa}}\\
  \displaystyle
  i\left(\frac{m}{2}\right)^{i\kappa}\frac{\Gamma(1/2- i\kappa)}
  {(-x_-\pm i\sigma)^{1/2- i\kappa}}+
  i\left(\frac{m}{2}\right)^{- i\kappa+1}\frac{\Gamma(-1/2+ i\kappa)}
  {(x_+\mp i\sigma)^{-1/2+ i\kappa}}
 \end{pmatrix}.
\end{equation}
So we see that at the surface of the light cone
$\psi_\kappa^{(\pm)}(x)$ contains $\delta(\kappa- i/2)$ at $x_+=0$
in the upper, and $\delta(\kappa+ i/2)$ at $x_-=0$ in the lower
component.

It was shown in the preceding section that in the case of a scalar
field the family of boost modes does not constitute a complete set
in MS after excluding the zero mode. This is because the scalar
boost modes possess a $\delta$-function singularity at the surface
of the light cone, and hence the point $\kappa=0$ gives finite
contribution to physical quantities. Some of them, e.g., Wightman
function, are determined by the spectral point $\kappa=0$ entirely.
There is a question whether it is possible to delete the point $\kappa=0$
from the spectrum in the fermion case. To clarify this issue we will consider the
integral
\begin{equation}\label{fk_d}
  \widetilde{\mathfrak{F}}^{(\pm)}(\mathfrak{h},0)=
  \lim\limits_{\delta\to0}\left(\int
  \limits_{-\infty}^{\kappa_0-\delta}+\int\limits_{\kappa_0+\delta}^\infty\right)
  d\kappa\,\psi^{(\pm)}_\kappa(0)\mathfrak{h}(\kappa)=\mathfrak{F}^{(\pm)}
  (\mathfrak{h},0)-  \lim\limits_{\delta\to0}\int\limits_{\kappa_0-\delta}^{\kappa_0+\delta}
  d\kappa\,\psi^{(\pm)}_\kappa(0)\mathfrak{h}(\kappa)\,,
\end{equation}
where $\kappa_0$ is an arbitrary real number.

Consider the upper component of the second term in the RHS of Eq.~(\ref{fk_d})
\begin{equation}\label{int0}
\frac{1}{2\pi}\lim\limits_{\delta\to0}
\lim\limits_{\sigma\to0}\int\limits_{\kappa_0-\delta}^{\kappa_0+\delta}
    d\kappa\,\left(\frac{m}{2}\right)^{-i\kappa}
  \frac{\Gamma(1/2+i\kappa)}{\sigma^{1/2+i\kappa}}\mathfrak{h}(\kappa).
\end{equation}
The integral in (\ref{int0}) can be easily calculated in the limit $\sigma\to0$ and we obtain
\begin{equation}\label{int01}
\begin{split}
\lim\limits_{\delta\to0}\int\limits_{\kappa_0-\delta}^{\kappa_0+\delta}
d\kappa\,[\psi^{(\pm)}_\kappa(0)]_{up}\mathfrak{h}(\kappa)=\frac{i}{2\pi}
\left(\frac{m}{2}\right)^{-i\kappa_0}\lim\limits_{\delta\to0}
\lim\limits_{\sigma\to0}\frac{\sigma^{-i\kappa_0-1/2}}{\ln\sigma}
\left[\left(\frac{m}{2}\right)^{-i\delta}\Gamma(1/2+i\kappa_0+i\delta)
\mathfrak{h}(\kappa_0+\delta)\sigma^{-i\delta}-\right.\\
\left.-\left
(\frac{m}{2}\right)^{i\delta}\Gamma(1/2+i\kappa_0-i\delta)
\mathfrak{h}(\kappa_0-\delta)\sigma^{i\delta}+O\left(\frac{1}{\ln\sigma}\right)
\right]
\end{split}
\end{equation}
Eq.~(\ref{int01}) evidently shows that the result of calculation for
$\widetilde{\mathfrak{F}}^{(\pm)}_{up}(\mathfrak{h},0)$ strongly
depends on the sequence of limit processing $\delta\to0$ and
$\sigma\to0$. If we changed the sequence of limit processing in
(\ref{int0}), (\ref{int01}), we would return from
$\widetilde{\mathfrak{F}}^{(\pm)}(\mathfrak{h},0)$ (\ref{fk_d}) to
${\mathfrak{F}}^{(\pm)}(\mathfrak{h},0)$ (\ref{fk}). Not
surprisingly, Eq.~(\ref{fk_d}) reproduces the result (\ref{defdel})
in this case. The accepted sequence which corresponds to calculation
of the functional (\ref{fk_d}) leads to a meaningless result. This
means that pricking of \emph{any} point out of the real axis $\kappa$ is
inadmissible in the fermion case. In other words, the set of boost
modes (\ref{fermbm}),(\ref{fermbm2}) does not constitute a complete
set in MS after deleting a single \emph{arbitrary} point from the
spectrum. It can be complete only in the space which is MS without
the line cone (four separate wedges of MS). In this sense our
conclusion repeats the analogous result for boson field discussed in
the preceding section.

It is worth noting that the latter result was obtained under minimum
possible restrictions as regards to the test functions
$\mathfrak{g}(p)$. If we restrict ourselves to $\mathfrak{g}(p)$
belonging to the class $K$ of smooth functions with compact support,
as it is often done in quantum field theory, functions
$\mathfrak{h}(\kappa)$ would belong to the class $Z$ \cite{gelfand}.
In this case the proof of impossibility of deleting an arbitrary
point $\kappa_0$ from the spectrum is especially simple. Indeed, the
Gel'fand's $\delta$-function is an analytic distribution in the
whole complex plain $\kappa$. Thus it can be expanded in a Taylor
series around any point $\kappa-\kappa_0$,
\begin{equation}\label{series}
\delta\left(\kappa\mp \frac{i}{2}\right)= \sum\limits_{n=0}^\infty\frac{1}{n!}
\left(\kappa_0\mp\frac{i}{2}\right)^n\delta^{(n)}(\kappa-\kappa_0)\,,
\end{equation}
and the radius of convergence of the series (\ref{series}) is equal
to infinity \cite{gelfand}. Here $\delta^{(n)}(\kappa-\kappa_0)$
denotes the $n$-th derivative of the $\delta$-function at the point
$\kappa-\kappa_0$. As a result,
\begin{equation}\label{PI}
\lim\limits_{\delta\to0}\left(\int
  \limits_{-\infty}^{\kappa_0-\delta}+\int\limits_{\kappa_0+\delta}^\infty\right)
  d\kappa\,\delta\left(\kappa\mp \frac{i}{2}\right)\mathfrak{h}(\kappa)
  =\sum\limits_{n=0}^\infty\frac{1}{n!}
\left(\kappa_0\mp\frac{i}{2}\right)^n\lim\limits_{\delta\to0}\left(\int
  \limits_{-\infty}^{\kappa_0-\delta}+\int\limits_{\kappa_0+\delta}^\infty\right)
  d\kappa\delta^{(n)}(\kappa-\kappa_0)\mathfrak{h}(\kappa)=0\,,
\end{equation}
and hence, on the class $Z$ of test functions, deleting of an
arbitrary spectral point $\kappa_0$ from the spectrum leads to
vanishing of functionals $\mathfrak{F}^{(\pm)}(\mathfrak{h},0)$
(\ref{kont_1}) at the vertex of the light cone, and its finite
variation at other points of the cone surface. However, since the
Gel'fand's $\delta$-function is analytic on the whole plane of
complex $\kappa$, it can be expanded in a Taylor series around
another real point $\kappa-\kappa_1$, $\kappa_1\neq\kappa_0$. If we
calculate integral (\ref{PI}) using such representation of the
$\delta$-function, we will see that deleting of the spectral point
$\kappa_0$ from the spectrum does not influence the result of
integration. Thus, since the value of the matrix element (\ref{kont_1})
depends on the way of calculation, we conclude that the operation of
exclusion of an arbitrary point from the spectrum is meaningless in
agreement with the previous consideration, see discussion of
Eq.~(\ref{fk_d}) in the preceding paragraph.

As it was shown in the preceding section, Wightman functions
$\Delta^{(\pm)}(x)$ of a massive scalar field are determined by the
zero boost modes $\Psi^{(\pm)}_0(x)$, see Eq.~(\ref{psi01}). This is
a consequence of translational invariance of $\Delta^{(\pm)}(x)$ and
the presence of Dirac $\delta$-function singularity of
$\Psi^{(\pm)}_{\kappa}(x)$ at the light cone. A similar result is
valid also for the Wightman functions of a massive fermion field.
Indeed, due to the property of translational invariance we can
write, compare (\ref{fpolnota}),
\begin{equation}\label{wf_f}
  S^{(\pm)}(x',x'')=i\int\limits_{-\infty}^\infty
  d\kappa\psi^{(\pm)}_\kappa(x)\psi^{(\pm)^\dag}_\kappa(0)\gamma^0\,,
  \quad x=x'-x''\,.
\end{equation}
Using now Eq.~(\ref{fermdelta}) we get the following result for
matrix elements $S^{(\pm)}_{\alpha\beta}(x)$ of Wightman function
(\ref{wf_f}),
\begin{equation}\label{mxelfw}
S^{(\pm)}_{\alpha1}(x)=\pm i\sqrt{\frac{m}{2}}\psi^{(\pm)}_
{\frac{i}{2},\alpha}(x)\,,\quad S^{(\pm)}_{\alpha2}(x)=
i\sqrt{\frac{m}{2}}\psi^{(\pm)}_{-\frac{i}{2},\alpha}(x)\,,\quad
\alpha=1,2\,,\end{equation} where $\psi^{(\pm)}_{\kappa,\alpha}(x)$
are the $\alpha$-components of the boost modes (\ref{fermbm}),
(\ref{fermbm2}). Thereby, we see that the matrix elements of
$S^{(\pm)}(x)$ are determined by only two "spectral points"
$\,\kappa=\pm i/2$. Exactly as in the scalar case, it is easy to
ascertain that this result holds valid for the smeared Wightman
functions as well.

Using Eq.~(\ref{fermbm2}) one can express the Wightman functions
(\ref{mxelfw}) in the form, compare (\ref{Wf_s}),
\begin{equation}\label{wffK}
\begin{array}{c}
  S^{(\pm)}(x)=\displaystyle i\frac{m}{2\pi}
  \begin{pmatrix}\pm K_0(w_{\pm})&\displaystyle\left(\frac{x_+\mp
  i\sigma}{x_-\mp i\sigma}
\right)^{-\frac{1}{2}}
K_1(w_{\pm})\\\displaystyle\left(\frac{x_+\mp
i\sigma}{x_-\mp i\sigma}\right)^{\frac{1}{2}}
K_1(w_{\pm})&\pm K_0(w_{\pm})\end{pmatrix}\,,
\end{array}
\end{equation}
where $w_{\pm}$ is the same as in Eq.~(\ref{scbm3}).

Consider now the case of a massless fermion field first studied in
Ref.~\cite{jetpl}. The Wightman function of the massless field
$S_0^{(\pm)}(x)$ can be obtained easily by passage to the limit
$m\to0$ in Eq.~(\ref{wffK})
\begin{equation}\label{wfm0}
  S^{(\pm)}_0(x)=\lim\limits_{m\to
  0}S^{(\pm)}(x)=\pm\frac{1}{2\pi}\begin{pmatrix}0&\displaystyle\frac{1}{x_+\mp
  i\sigma}\\\displaystyle\frac{1}{x_-\mp
  i\sigma}&0\end{pmatrix},\quad\sigma\to0.
\end{equation}

Just as in the cases considered earlier, one can use
$S^{(\pm)}_0(x-\alpha_q)$ to obtain any complete set of orthonormalized
solutions of Dirac equation. In particular, for the boost modes we
have,
\begin{equation}\label{m0razl}
  \chi^{(\pm)}_\kappa(x)=\int\limits_{-\infty}^\infty dq
  S_0^{(\pm)}(x-\alpha_q)e^{-i\kappa q-\frac{q}{2}\sigma_3}C,
\end{equation}
where $C=\begin{pmatrix}c_1\\c_2\end{pmatrix}$ is an arbitrary
two-component column, compare (\ref{fermrazl}), (\ref{f_k}). We
see that at $m=0$ the Wightman function  (\ref{wfm0}) is antidiagonal.
Owing to this property the upper and the lower components of the
boost mode (\ref{m0razl}) become independent:
\begin{equation}\label{ppm}
 \chi^{(\pm)}_\kappa(x)=\tilde{c}_1\chi^{(\pm)}_{\kappa+}(x)e_++\tilde{c}_2
\chi^{(\pm)}_{\kappa-}(x)e_-,
\end{equation}
where $e_{\pm}$ are the eigenvectors of the Pauli matrix $\sigma_3$,
$\sigma_3 e_{\pm}=\pm e_{\pm}$. This is because there appears a new
conservation law for the massless case, conservation of chirality,
see, e.g., \cite{chir}. Thereby solutions of the Dirac equation
(\ref{direq}) are labeled by the additional quantum number
$\tau=\pm\,$, chirality. The coefficients $\tilde{c}_1$ and $\tilde{c}_2$ in
(\ref{ppm}) are independent and can be determined by the
normalization condition for solutions $\chi^{(\pm)}_{\kappa+}(x)e_+$
and $\chi^{(\pm)}_{\kappa-}(x)e_-$ separately. Note that in
the massive case, since chirality is not conserved, the components
of the column $C$ (\ref{f_k}) cannot be determined separately. In
that case their combination forms a monomial factor of the solution
$\psi^{(\pm)}_\kappa(x)$ (\ref{fermbmgen}).

For the normalized positive frequency functions $\chi^{(\pm)}_{\kappa\tau}$ we have
\begin{equation}\label{m0boost}
  \chi^{(\pm)}_{\kappa\tau}(x)=\frac{e^{\mp i\frac{\pi}{2}\lambda}\Gamma
  (\lambda)}{2\pi(x_\tau\mp
  i\sigma)^\lambda},\quad\lambda=\frac{1}{2}+i\tau\kappa,\quad\sigma\to 0,
\end{equation}
where the phase factors were chosen for the sake of convenience. It
is worth noting that the negative frequency functions
$\chi^{(-)}_{\kappa\tau}$ come out of (\ref{m0boost}) by complex
conjugation and the change $\kappa\rightarrow -\kappa$.

Plane waves are also labeled by the quantum number $\tau$ in the
massless case,
\begin{equation}\label{m0plane}
\begin{array}{c}\displaystyle
  \varphi^{(\pm)}_{p\tau}(x)=\displaystyle\frac{1}{\sqrt{2\pi}}e^{\mp
  ipx_\tau},\quad 0<p<\infty \,.
\end{array}
\end{equation}
They are linked to the boost modes
$\chi^{(\pm)}_{\kappa\tau}(x)e_\tau$ through the Mellin transform
\begin{equation}\label{mellin}
  \chi^{(\pm)}_{\kappa\tau}(x_\tau)=\int
  \limits_0^\infty\frac{dp}{\sqrt{2\pi}}p^{\lambda-1}
  \varphi^{(\pm)}_{p\tau}(x_\tau).
\end{equation}
This means that, if we assume that distributions
$\varphi^{(\pm)}_{p\tau}(x_\tau)$ are defined on the same class of
test functions $\mathfrak{g}(p)$ as in the massive case, the
distributions $\chi^{(\pm)}_{\kappa\tau}(x_\tau)$ will be defined on
the class of test functions $\mathfrak{h}(\kappa)$ analytic in the
strip (\ref{polosa}) and descending at $|\kappa|\to\infty$ in this
strip.

Then taking into account Eq.~(\ref{delta}) we conclude that boost
modes of a massless fermion field at the surface of the light cone
are $\delta$-functions of a complex argument, compare \cite{jetpl},
\begin{equation}\label{phichicone}
  \chi_{\kappa\tau}^{(\pm)}(x_\tau=0)=\delta\left(\kappa-\tau \frac{i}{2}\right)\,.
\end{equation}

Now we will show how the modes (\ref{m0boost}) can be obtained from
Eq.~(\ref{fermbm2}) by limit processing $m\rightarrow 0$. Using the
ascending series \cite{GrR} for functions $K_{1/2\pm
i\kappa}(w_{\pm})$ we get
\begin{equation}
\displaystyle \label{psilim}
 \psi_\kappa^{(\pm)}(x)= \left(\frac{m}{2}\right)^{-i\kappa}\chi_{\kappa+}^{(\pm)}\,e_+\pm
\left(\frac{m}{2}\right)^{i\kappa}\chi_{\kappa-}^{(\pm)}\,e_-,
\end{equation}
i.e., representation (\ref{ppm}) with coefficients $\tilde{c}_1$ and
$\tilde{c}_2$ containing singular at $m\rightarrow0$ phase factors.
However these factors do not influence the normalization constants,
have no impact on any physical quantities and hence can be omitted.

\section{Concluding remarks}

We have shown that Wightman function of a free quantum field
generates any complete set of solutions of relativistic wave
equations. Using this approach we have constructed the complete sets
of solutions to KFG and Dirac equations consisting of eigenfunctions
of the generator of Lorentz rotations (boost operator).

Boost modes are used as a basis for field quantization very rarely.
Till now they were exploited only for analysis of the so-called
"Unruh effect" \cite{unruh,narozhny} and at attempts to quantize a
charged massive scalar field in the presence of an external constant
electric field \cite{gabriel,physlett}. However, there are many
problems, especially in the quantum field theory in a curved space,
where the boost symmetry may appear to be the only symmetry for the
quantum field, and thus using it for separation of variables in a
classical field equation is the only instrument to find solutions
for such equations.

The specific feature of the boost modes is that, taken at the
surface of the light cone, they as functions of the boost quantum
number $\kappa$ possess strong singularities. Certainly, the
physical reason for these singularities is the singularity of
Lorentz transformations at $v=c$.

For the case of a scalar field this is a Dirac $\delta$-function
singularity $\Psi_\kappa(x)\big|_{x_{\pm}=0}\sim\delta(\kappa)$
\cite{narozhny}. This leads to a special role of the zero boost
mode: its exclusion from the set of the boost modes makes the latter
incomplete. The exceptional role of the zero boost mode becomes
quite clear if we recall that the zero value of $\kappa$ means that
$\Psi_0(x)$ is a Lorentz invariant positive frequency solution of
KFG equation, i.e. the Wightman function for the quantum field
coinciding with the positive frequency part of the commutator of two
scalar field operators. Hence the exclusion of the zero boost mode
results in a "quantum"$\,$ theory with commuting field operators.

The singularities of the fermion boost modes are even stronger. It
is shown in the present paper that at the surface of the light cone
they possess $\delta$-function of a complex argument
$\psi_{\kappa\alpha}(x)\big|_{x_{\pm}=0}\sim\delta(\kappa\mp i/2)$.
The $\delta$-function of a complex argument was first introduced by
Gel'fand and Shilov in Ref.~\cite{gelfand}, and was defined on the
class $Z$ of entire functions. Our $\delta$-function is defined on
the class of test functions analytic in the strip (\ref{polosa}).
Actually, the width of the strip is determined by physical
requirements. Our choice provides square integrability of
one-particle wave packets and finiteness of their energy. If we
require finiteness of the squared energy we should narrow the class
of test functions and extend the width of the strip to
$$-3/2-\epsilon<\rm{Im}\kappa<3/2+\epsilon\,,\quad \epsilon>0\,.$$
Further toughening of requirements to physical states will lead to
subsequent extension of the width of the strip of analyticity of the
test functions. Therefore it is reasonable to generalize the concept
of $\delta$-function of a complex argument and introduce the
distribution $\delta_s(\kappa)$
\begin{equation}\label{delta_s}
(\delta_s(\kappa-\kappa_0),f(\kappa))=f(\kappa_0)\,,
\end{equation}
defined on the class $Z_s$ of test functions $f(\kappa)$ analytic in
the strip
\begin{equation}\label{polosa_s}
-s-\epsilon<{\rm Im}\kappa<s+\epsilon\,,\quad s>0,\, \epsilon>0\,,
\end{equation}
and $\kappa_0$ belong to this strip. In such notation the
$\delta$-function introduced in Sec. III will look as
$\delta_{1/2}(\kappa)$, the Gel'fand $\delta$-function as
$\delta_{\infty}(\kappa)$ and the standard Dirac $\delta$-function
as $\delta_0(\kappa)$.

It is worth noting that for $\delta$-function defined on the class
$Z_s$ the following integral representation is valid
\begin{equation}
 \delta_s(\nu)=\frac{1}{2\pi}\int\limits_{-\infty}^\infty dq
 e^{-i\nu q},\quad\nu=\kappa\mp is,\quad s>0,
\end{equation}
which at $s=0$ is a straightforward generalization of the standard
representation for the Dirac $\delta$-function. Besides, the
following relations are valid,
\begin{equation}
\delta_s(\nu)=\frac{i}{2}H_{i\nu}^{(1)}(0)=-\frac{i}{2}H_{i\nu}^{(2)}(0)
=\frac{1}{\pi}K_{i\nu}(0),\quad\nu=\kappa\mp
is,\quad s>0,
\end{equation}
compare Ref.~\cite{lvov} for the case $s=0$.

The presence of a $\delta$-function of a complex argument in a boost
mode at the surface of the light cone does not allow to exclude
\emph{any} point from the boost spectrum. We have shown this for the
case $s=1/2$, see Eqs.~(\ref{fk_d})-(\ref{PI}), and this statement
proving does not change for an arbitrary value of $s$. Another way
to prove this statement was used for the distribution
$\delta_{\infty}(\kappa)$ based on its analyticity in the whole
complex plane. The analogous proof could be given for the
$\delta_s$-function as well. In this case, due to the finite value
of radius of convergence $R=s+\epsilon$, the expansion
(\ref{series}) can be applied only for real
$\kappa_0=b_1\,,\,\,|b_1|<\epsilon$. As a next step the Dirac
$\delta$-function $\delta_0(\kappa-b_1)$ and every its derivative
can be represented as a Taylor series of the type (\ref{series})
centered at the point $b_2$ on the real axis,
$|b_2-b_1|<s+\epsilon$. So, $\delta_{s}(\kappa-is)$ will be expanded
in a $2$-multiple series of the Dirac
$\delta_0(\kappa-b_2)$-function and its derivatives. After a finite
number of steps $N$ we can reach an arbitrary spectral point
$\kappa_0$ and hence obtain a representation of
$\delta_{s}(\kappa-is)$ in the form of a $N$-multiple series of the
Dirac $\delta_0(\kappa-\kappa_0)$-function and its derivatives. It
is clear that the point $\kappa_0$ cannot be excluded from the
spectrum then.

The latter reasoning clearly explains why the scalar case, when we
have the only one distinguished point $\kappa=0$ which cannot be
excluded from the spectrum, differs drastically from the fermion
case. Indeed, the strip of analyticity (\ref{polosa_s}) for
$\delta_s(\kappa)$ degenerates into the real axis of the $\kappa$
complex plane at $s=0$, so that the radius of convergence for the
corresponding Taylor series becomes equal to zero, or in other words
the Dirac $\delta_0(\kappa)$-function is not an analytic
distribution. Thus the procedure discussed above cannot be realized.

To conclude, it is worth emphasizing that we have shown explicitly
that smearing of boost modes, or Wightman functions does not change
our results, see also Ref.~\cite{repl}.

\section{Acknowledgements}

This work was supported by Russian Fund for Basic Research, the
Ministry of Science and Education of the Russian Federation and the
Russian Federal Program "Scientific and scientific-pedagogical
personnel of innovative Russia".

\end{document}